\documentclass{emulateapj}
\usepackage{apjfonts}

\gdef\h50min{$h_{50}^{-1}$}
\gdef\h70min{$h_{70}^{-1}$}
\gdef\kms{km\,s$^{-1}$}

\gdef\msun{$M_{\odot}$}

\lefthead{}
\righthead{}
\slugcomment{Accepted for publication in Astrophysical Journal Letters}
\begin{document}

\title{The space density and colors of massive galaxies at $2<z<3$:
the predominance of Distant Red Galaxies}
\author{
P.~G.~van Dokkum\altaffilmark{1},
R.~Quadri\altaffilmark{1},
D.~Marchesini\altaffilmark{1},
G.~Rudnick\altaffilmark{2},
M.~Franx\altaffilmark{3},
E.~Gawiser\altaffilmark{1,4},
D.~Herrera\altaffilmark{1},
S.~Wuyts\altaffilmark{3},
P.~Lira\altaffilmark{4},
I.~Labb\'e\altaffilmark{5,6},
J.~Maza\altaffilmark{4},
G.~D.~Illingworth\altaffilmark{7},
N.~M.~F\"orster Schreiber\altaffilmark{8},
M.~Kriek\altaffilmark{3},
H.-W.~Rix\altaffilmark{9},
E.~N.~Taylor\altaffilmark{3},
S.~Toft\altaffilmark{1},
T.~Webb\altaffilmark{3}, and
S.~K.~Yi\altaffilmark{10}
}

\altaffiltext{1}{Department of Astronomy, Yale
University, New Haven, CT 06520-8101}
\altaffiltext{2}{National Optical Astronomical Observatory, 950 North
Cherry Avenue, Tucson, AZ 85721}
\altaffiltext{3}{Leiden Observatory, PO Box 9513, NL-2300 RA,
Leiden, The Netherlands}
\altaffiltext{4}{Departamento de Astronom\'\i{}a, Universidad de Chile,
Casilla 36-D, Santiago, Chile}
\altaffiltext{5}{Carnegie Observatories, 813 Santa Barbara Street, Pasadena,
CA 91101}
\altaffiltext{6}{Carnegie Fellow}
\altaffiltext{7}{UCO/Lick Observatory, University of California, Santa
Cruz, CA 95064}
\altaffiltext{8}{MPE, Giessenbachstrasse, Postfach 1312, D-85748
Garching, Germany}
\altaffiltext{9}{MPA, K\"o{}nigstuhl 17, D-69117 Heidelberg, Germany}
\altaffiltext{10}{Yonsei University, Seodaemoon-gu Shinchon-dong 134,
Seoul 120-749, South Korea}

\begin{abstract}

Using the deep multi-wavelength 
MUSYC, GOODS, and FIRES surveys
we construct a stellar mass-limited sample
of galaxies at $2<z<3$. The sample comprises 294 galaxies with
$M>10^{11}$\,\msun\ distributed over four independent fields
with a total area of almost 400\,arcmin$^2$.
The mean number density of massive galaxies in this redshift
range $\rho (M>10^{11}\,M_{\odot}) = (2.2 \pm 0.6)
\times 10^{-4}\,h_{70}^3$\,Mpc$^{-3}$. We present median values and
25$^{\rm th}$ and 75$^{\rm th}$ percentiles for the distributions of
observed $R_{\rm AB}$ magnitudes, observed
$J-K_s$ colors, and rest-frame ultra-violet continuum slopes,
$M/L_V$ ratios, and $U-V$ colors. The galaxies show a large
range in all these properties. The ``median galaxy''  is
faint in the observer's optical ($R_{\rm AB}=25.9$),
red in the observed near-IR ($J-K_s = 2.48$), has a rest-frame
UV spectrum which is relatively flat in $F_{\lambda}$ ($\beta= -0.4$),
and rest-frame optical colors resembling those of nearby spiral galaxies
($U-V = 0.62$). We determine which galaxies would be selected as Lyman break
galaxies (LBGs) or Distant Red Galaxies (DRGs, having $J-K_s>2.3$) in this
mass-limited sample. By number DRGs make up 69\,\% of the sample
and LBGs 20\,\%, with a small amount of overlap. By mass
DRGs make up 77\,\% and LBGs 17\,\%. Neither technique
provides a representative sample
of massive galaxies at $2<z<3$ as they only sample the extremes
of the population. As we show here, multi-wavelength
surveys with high quality photometry are essential for an unbiased census
of massive galaxies in the early Universe. The main uncertainty
in this analysis is our reliance on photometric redshifts;
confirmation of the results presented here requires extensive
near-infrared spectroscopy of optically-faint samples.

\end{abstract}


\section{Introduction}

The properties of massive galaxies at high redshift
place important constraints on galaxy formation models
(see, e.g., {Kauffmann} \& {Charlot} 1998; {Nagamine} {et~al.} 2005).
The ``standard'' and most successful method for
finding distant galaxies is the Lyman dropout technique, which relies on
the strong break in the rest-frame ultra-violet (UV) spectra of high
redshift galaxies blueward of the Lyman limit ({Steidel} {et~al.} 1996, 1999).
However, it is not yet clear whether these galaxies
are representative of the high redshift galaxy population,
in particular at the high-mass end.
As the Lyman break selection requires that galaxies
are very bright in the rest-frame UV it may miss objects that are
heavily obscured by dust or whose light is dominated by evolved
stellar populations.

Advances in instrumentation have made it
possible to select galaxies in complementary ways, 
and recent studies have demonstrated that the
universe at $z>2$ is much more diverse than had been realized.
Among recently identified ``new'' galaxy populations
are submm galaxies (e.g., Smail et al.\ 2004),
distant red galaxies (DRGs) selected by the criterion
$J-K_s>2.3$ ({Franx} {et~al.} 2003; {van Dokkum} {et~al.} 2003),
``IRAC Extremely Red Objects'' (IEROs; {Yan} {et~al.} 2004),
and ``$BzK$'' objects (Daddi et al.\ 2004).

The current situation is somewhat confusing, as the relative
contributions of the various newly identified galaxy populations
to the stellar mass budget and the cosmic star formation rate
are still unclear. Furthermore,
as emphasized by, e.g., {Adelberger} {et~al.} (2005) and Reddy et al.\ (2005),
there can be considerable overlap between selection
techniques, which makes it difficult to interpret the plethora of
windows on the high redshift Universe.

Ideally samples of
high redshift galaxies are selected not
by color or luminosity
but by stellar mass.
Whereas  luminosities and colors can vary
dramatically due to star bursts and the presence of dust
the mass evolution of galaxies is probably gradual.
Also, galaxy formation models can predict masses with somewhat higher
confidence than luminosities and colors.
Stellar masses of distant galaxies are usually determined by
fitting stellar population synthesis models to
broad band photometry. Although there are significant
systematic uncertainties, the stellar mass
of a galaxy is usually better constrained than
the instantaneous star formation rate,
age, or dust content (e.g., {Shapley} {et~al.} 2001; {Papovich} {et~al.} 2001; {van Dokkum} {et~al.} 2004; {F{\" o}rster Schreiber} {et~al.} 2004).

In this Letter we explore the properties of a stellar-mass limited
sample of galaxies. The main purpose is to measure ``basic''
aspects of massive galaxies to
compare with simulations of galaxy formation: their density
and colors.
A secondary goal is to quantify selection
biases introduced by two of the
most widely used techniques for identifying distant galaxies: the
Lyman break technique and the $J-K$ color selection of
Franx et al.\ (2003). We assume $\Omega_m=0.3$, $\Omega_{\Lambda}
=0.7$, and $H_0 = 70$\,\kms\,Mpc$^{-1}$. All magnitudes are
on the Vega system, unless identified as ``AB''.

\begin{figure}[t]
\epsscale{1.1}
\plotone{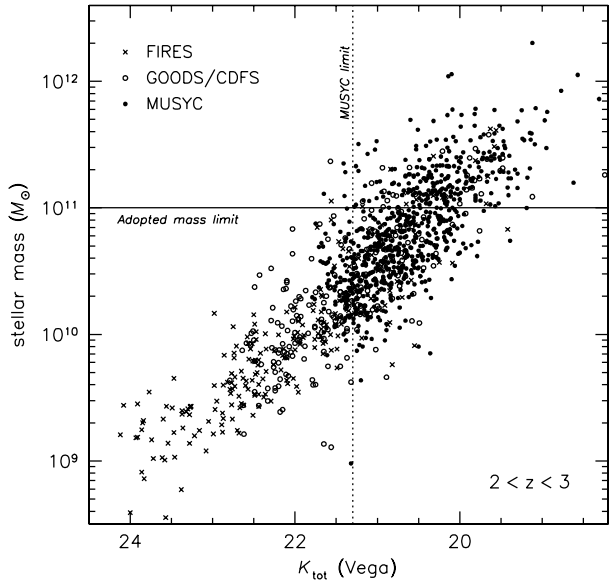}
\caption{
Relation between stellar mass and observed total $K_s$ magnitude for
galaxies at $2<z_{\rm phot}<3$.
The solid line shows our selection limit
of $M>10^{11}$\,\msun. The broken line shows the approximate photometric
limit of the MUSYC survey, which accounts for the majority of galaxies in our
sample. There are very few galaxies in FIRES and GOODS
with $M>10^{11}$\,\msun\ and
$K_s>21.3$, and we estimate
that the completeness
of the full sample of $M>10^{11}$\,\msun\ galaxies is
$\approx 95\,\%$.
\label{kmass.plot}}
\end{figure}

\section{Definition of the sample}

The sample is drawn from three deep multi-wavelength surveys, all having
high-quality optical -- near-IR photometry: the ``ultra-deep''
Faint InfraRed Extagalactic Survey (FIRES; Franx et al.\ 2003),
the Great Observatories Origins Deep Survey (GOODS; Giavalisco et al.\
2004) Chandra Deep Field South (CDFS), and
the new Multi-wavelength Survey by Yale-Chile (MUSYC;
Gawiser et al.\ 2006; Quadri et al.\ in prep).

Photometric catalogs were
created for all fields in the same way, following the procedures
of {Labb{\' e}} {et~al.} (2003).
Photometric redshifts were determined using the method of
{Rudnick} {et~al.} (2001, 2003). Comparing the photometric
redshifts with 696 spectroscopic redshifts (63 at $z\geq 1.5$)
gives a scatter in $\Delta z / (1+z)$ of $\sigma
= 0.06$. Restricting the analysis to galaxies at $z\geq 1.5$
in the MUSYC fields gives $\sigma = 0.12$,
corresponding to $\Delta z \approx 0.4$ at $z=2.5$. Approximately
$5$\,\% of galaxies in this sample are ``catastrophic''
outliers. A full discussion of the quality of the
photometric redshifts is given elsewhere (G.\ Rudnick et al.,
in prep). To determine masses,
stellar population synthesis models were fit to the photometry
using standard techniques (see, e.g., {Shapley} {et~al.} 2001; {F{\" o}rster Schreiber} {et~al.} 2004). {Bruzual} \& {Charlot} (2003) models were used, with solar metallicity
({Shapley} {et~al.} 2004; {van Dokkum} {et~al.} 2004) and a {Salpeter} (1955) initial
mass function from $0.1$\,\msun\ to $100$\,\msun.
Star formation histories were parameterized by
a declining star formation rate with characteristic timescale
$\tau = 0.3$\,Gyr (see {F{\" o}rster Schreiber} {et~al.} 2004).
The {Calzetti} (1997) reddening law was used, with
extinction ranging from $A_V=0$ to $A_V=3$.
We note that the derived masses are probably not significantly
affected by the presence of AGN (e.g., Rubin et al.\ 2004,
Reddy et al.\ 2005), as their contributions to the
broad band fluxes are probably small (F\"orster Schreiber et al.\
2004; Webb et al.\ 2006).

Fig.\ \ref{kmass.plot} shows the relation between stellar mass and
observed total $K_s$ magnitude for galaxies with photometric redshifts
$2<z<3$. There is a clear relation, with a scatter of a factor
of $\sim 10$. 
We selected all 294 galaxies with $2<z_{\rm phot}<3$
and stellar masses $M>10^{11}$\,\msun\ in the three surveys.
The reliability of this procedure
was assessed in the following ways. First, we
compared the masses derived from $U-K_s$ photometry to
masses derived from $U-K_s$ plus Spitzer/IRAC photometry in
the HDFS (Labbe et al.\ 2005). Although the masses of
individual galaxies can vary by $\sim 30$\,\% the systematic difference
is $\lesssim 10$\,\%. Next, we determined what fraction of massive
galaxies are fainter than $K_s=21.3$, the approximate limit of
the MUSYC survey. Only $5$\,\% of
galaxies with $M>10^{11}$\,\msun\ in the deep FIRES and
GOODS fields have $K_s>21.3$.
Extremely obscured massive galaxies
could be missed even in the deep FIRES and GOODS data, but
the fact that
$\sim 90$\,\% of submm-selected galaxies at $z\sim 2.2$ have
$K<21$ (Smail et al.\ 2004) implies that such objects are very rare.
We conclude that our mass-limited sample
of 294 galaxies at $2<z<3$ is $\sim 95$\,\% complete.

\section{Density}

The FIRES, GOODS, and MUSYC surveys cover four independent fields:
FIRES MS\,1054--03 (23\,arcmin$^2$), 
GOODS CDFS\footnote{Area with $JHK_s$ coverage.} (69\,arcmin$^2$),
MUSYC SDSS\,1030 (103\,arcmin$^2$) and
MUSYC HDFS (188\,arcmin$^2$). The total area is 383\,arcmin$^2$,
of which 76\,\% is contributed by MUSYC.
The average
surface density of $M>10^{11}$\,\msun\ galaxies with $2<z<3$ is
0.71\,arcmin$^{-2}$, but there are large field-to-field variations.
The density in the CDFS field is only 0.42\,arcmin$^{-2}$,
60\,\% of the mean and a factor of three
lower than that of the highest density field, SDSS\,1030.
This large variation is indicative of strong clustering and implies that
densities inferred from individual $\sim 100$\,arcmin$^2$
fields should be treated with caution.

After a 5\,\% correction for incompleteness the mean space
density  $\rho(M>10^{11}\,M_{\odot}) =
(2.2 \pm 0.6) \times 10^{-4}$\,Mpc$^{-3}$. The uncertainty
includes the effects of field-to-field variations, but does not
include possible effects caused by systematic errors in the
photometric redshifts (see, e.g., Shapley et al.\ 2005).
We note
that this density is a factor of $\sim 5$ lower than that of
$z\approx 3$ $U$-dropout galaxies to $R_{\rm AB}=25.5$
({Steidel} {et~al.} 1999), which typically have much lower
masses.

\section{Properties of massive galaxies at $2<z<3$}

We use our mass-limited sample of 294 galaxies to determine
the median and dispersion in observed and rest-frame
properties of the galaxies. Table 1 gives the
median and 25/75-percentiles of the distributions of
observed $R_{\rm AB}$ magnitude and $J-K_s$ color;
rest-frame $U-V$ color and $M/L_V$ ratio; and
rest-frame UV slope, parameterized by $F_{\lambda}
\propto \lambda^{\beta}$.
The rest-frame $V$ magnitudes and
$U-V$ colors were determined from the observed
magnitudes following similar procedures as
those outlined in {van Dokkum} \& {Franx} (1996).
Rest-UV slopes $\beta$ were determined
from the best fitting spectral energy distributions (SEDs),
following the
{Calzetti}, {Kinney}, \&  {Storchi-Bergmann} (1994) method of fitting to
the ten rest-UV bins defined by those authors.

\begin{figure}[t]
\epsscale{1.1}
\plotone{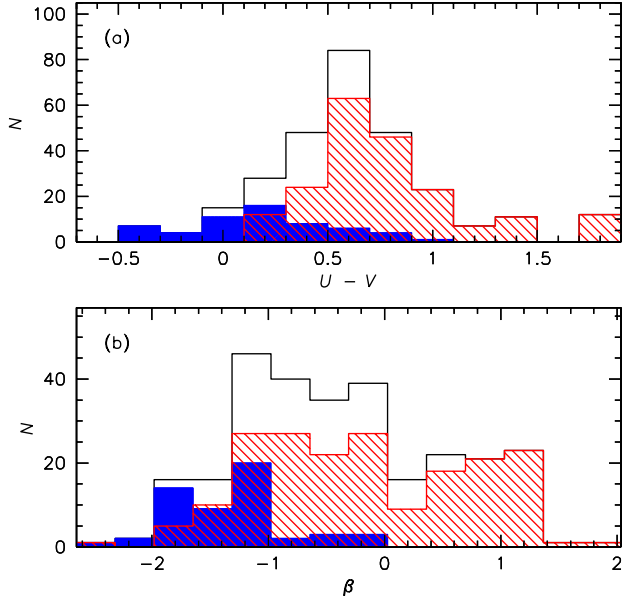}
\caption{
Distribution of rest-frame $U-V$ colors (a) and
rest-frame UV slope $\beta$ (b) for galaxies with
$M>10^{11}$\,\msun\ and $2<z_{\rm phot}<3$.
The galaxies show a wide range in optical- and
rest-UV colors. Blue histograms indicate galaxies
with the colors and luminosities of LBGs;
red histograms indicate DRGs.
\label{mluv.plot}}
\end{figure}

\begin{figure}[t]
\epsscale{1.1}
\plotone{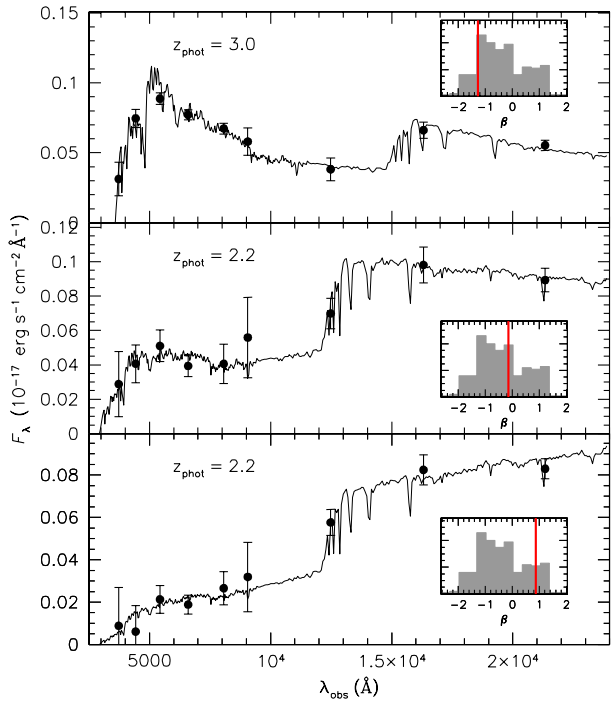}
\caption{
Spectral energy distributions of three MUSYC galaxies with different
rest-UV slope $\beta$. Overplotted are the best fitting Bruzual
\& Charlot (2003) models.
From top to bottom the galaxies have
$\beta=-1.3$, $-0.2$, and $0.9$ respectively. The blue rest-UV
SED of the top object is typical for LBGs. The slope of
the middle galaxy is close to the median value
of the full sample.
\label{seds.plot}}
\end{figure}

As can be inferred from
Table 1 the galaxies span a large range in all these properties.
The ``median galaxy'' is red and faint in the
observer's optical, with $\langle R_{\rm AB}\rangle = 25.9$.
We show the full distribution of the rest-frame $U-V$
colors in Fig.\ \ref{mluv.plot}(a).
The bluest galaxies have $U-V<-0.1$
and are bluer than nearby
irregular galaxies, whereas the colors of the reddest galaxies
are similar to those of nearby ellipticals
(see, e.g., {Fukugita}, {Shimasaku}, \&  {Ichikawa} 1995).
The median $\langle U-V\rangle = 0.6$, which is similar to
nearby spiral galaxies but also to nearby dust-enshrouded star
burst galaxies (e.g., {Armus}, {Heckman}, \& {Miley} 1989).

The distribution of $\beta$ is shown in Fig.\ \ref{mluv.plot}(b).
Remarkably, the distribution is rather flat and has no well-defined
peak, in contrast to previous studies of optically-selected
samples ({Adelberger} \& {Steidel} 2000). The median $\langle \beta \rangle
=-0.39$, indicating a relatively flat spectrum in $F_{\lambda}$.
A potential worry is that individual values of $\beta$ are uncertain,
as many galaxies are very faint in the observer's optical.
We tested the robustness of the derived distribution of $\beta$ by
summing the observed optical fluxes of the galaxies in the
lower and upper 25\,\% quartiles,
weighting by the inverse of the total optical flux. The
power-law slopes of these summed SEDs are in very good
agreement with the median $\beta$s that we determined
from the SED fits.

The large range of properties of massive galaxies at $2<z<3$
is illustrated in Fig.\ \ref{seds.plot}, which shows the
full $UBVRIzJHK_s$ SEDs
of three galaxies from the MUSYC
survey with different values of $\beta$. The top galaxy has
a very blue SED similar to those of UV-selected samples
(see, e.g., {Shapley} {et~al.} 2001), the middle object has
an SED which resembles that of nearby spiral galaxies,
and the bottom galaxy
has a very red SED indicating strong extinction.


\begin{small}
\begin{center}
{ {\sc TABLE 1} \\
\sc Observed and Rest-frame Properties} \\
\vspace{0.1cm}
\begin{tabular}{lrrr}
\hline
\hline
Quantity & 25\,\% & Median & 75\,\% \\
\hline
$R_{\rm tot, AB}$ (obs) & 25.1 & 25.9 & 26.7 \\
$J-K_s$ (obs) & 2.22 & 2.48 & 2.85 \\
$U-V$ (rest) & 0.41 & 0.62 & 0.80 \\
$\log M/L_V$ (rest) & 0.05 & 0.20 & 0.30 \\
$\beta$ (rest) & $-1.13$ & $-0.39$ & 0.31 \\
\hline
\end{tabular}
\end{center}
\end{small}

\section{Discussion}

The main result of our analysis is that massive galaxies at $z\sim 2.5$
span a large range in rest-frame UV slopes, rest-frame optical colors,
and rest-frame $M/L_V$ ratios, indicating significant variation in
dust content, star formation histories, or both. This result is not
surprising in the
light of the recent discoveries of DRGs, IEROs, and other populations.
Here we have quantified the median colors and their range for
a uniformly selected, large, mass-limited sample.

\begin{figure}[t]
\epsscale{1.1}
\plotone{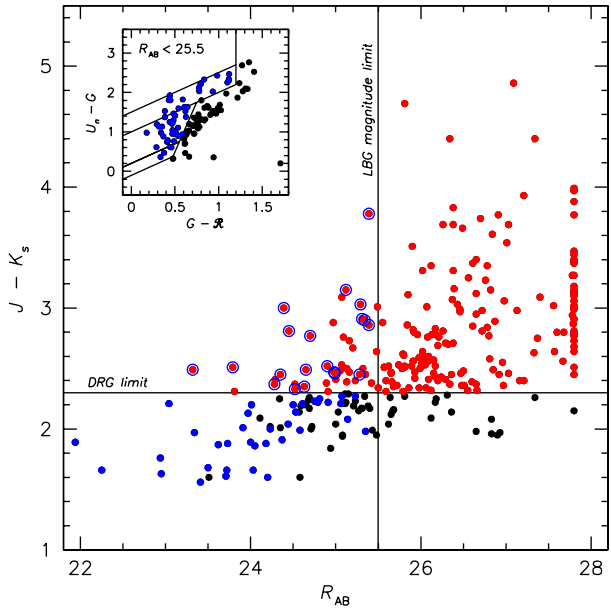}
\caption{
Correlation between observed $J-K_s$ color and $R_{\rm AB}$
magnitude. The majority
of the galaxies are faint in $R$ and red in $J-K_s$. Red symbols denote
DRGs with $J-K_s>2.3$; blue points denote LBGs with $R_{\rm AB}<25.5$.
Red symbols with blue circles fall in both categories.
Massive LBGs have blue near-IR colors and are bright in the observed
$R$ band. The inset shows the optical color distribution of galaxies
with $R<25.5$. Only $\sim 50$\,\% of optically-bright massive
galaxies have the colors of LBGs.
\label{jkr.plot}}
\end{figure}

The large variation in the rest-frame color distributions of
our mass-limited sample implies that ``standard'' color selection
techniques produce biased samples. We consider
two of the two most widely
used selection techniques in this redshift range: the
Lyman break technique of Steidel and collaborators and the
$J-K_s>2.3$ Distant Red Galaxy (DRG) selection
({Franx} {et~al.} 2003; van Dokkum et al.\ 2003). Lyman break galaxies are
identified in the following way. From
the best-fitting {Bruzual} \& {Charlot} (2003) SEDs (which include
absorption due
to the Ly\,$\alpha$ forest) we
calculated synthetic colors in Steidel's
$U_n G \cal R$ system. To qualify as an
LBG an object has to have $R_{\rm AB}<25.5$ and synthetic $U_n G \cal R$
colors which place it in the Lyman break, BX, or BM selection
region (see {Steidel} {et~al.} 2003, 2004).
Combined these criteria provide a continuous
selection of  galaxies over the redshift range considered
here.\footnote{A LBG in this definition is therefore an object
which has $R_{\rm AB}<25.5$,
$2<z_{\rm phot}<3$, and
is either a classical ``U-dropout'' or a BX/BM object.}
Figure \ref{jkr.plot} illustrates the LBG and DRG selection
techniques, as applied to our sample. DRGs with $J-K_s>2.3$
are indicated by red symbols and LBGs by blue symbols. The
DRG limit and the ``standard'' photometric LBG limit of $R_{\rm AB}
=25.5$ are also indicated.\footnote{We note that not all galaxies
with $J-K_s>2.3$ have redshifts in the range $2<z<3$: to $K_s=21$ we
find that $\sim 50$\,\%
are in this redshift range with the rest
about equally split between $z<2$ and $z>3$ galaxies.}

By number, DRGs make up 69\,\% of the sample and
LBGs 20\,\%. The DRG and LBG samples do not show much overlap:
only 7\,\% of objects fall in both
categories. By rest-frame $V$-band
luminosity DRGs contribute 64\,\% and LBGs 32\,\%. By mass
DRGs contribute 77\,\% and LBGs 17\,\%.
Together, the LBG and DRG techniques
identify 82\,\% of massive galaxies by number and
84\,\% by mass. Most of the remaining
galaxies are optically faint, slightly bluer than the
$J-K_s=2.3$ limit, and have redshifts $z<2.5$.
Approximately 85\,\% of them fall in the ``$BzK$''
selection region (Daddi et al.\ 2004), which is optimized
for galaxies at $1.4<z<2.5$.
We note that the relatively small fraction
of LBGs in the sample is not solely due to the
imposed $R_{\rm AB}<25.5$ limit.
As shown in the inset of
Fig.\ \ref{jkr.plot} only $\sim 50$\,\% of galaxies with
$R_{\rm AB}<25.5$ have the rest-UV colors of LBGs, and
this fraction decreases going to
fainter $R$ magnitudes: when no $R$ limit is imposed
we find that $\sim 1/3$ of the galaxies have the
colors of LBGs. The underlying reason is the
broad distribution of $\beta$.

It is clear from Fig.\ \ref{jkr.plot} that
the LBG selection produces very different samples of
massive high redshift galaxies than the DRG selection.
Both samples are biased: LBGs are too blue and DRGs are too red
when compared to the median values of the full sample.
This bias is shown explicitly by the blue and red histograms
in Fig.\ \ref{mluv.plot}.
%
The Lyman break criteria were designed to find star forming
galaxies, but
{Shapley} {et~al.} (2004) and {Adelberger} {et~al.} (2005) have argued
that they can also be used to find
massive galaxies at high redshift. This is obviously the
case, but we find that the colors and $M/L_V$ ratios of
massive LBGs are not representative for the full sample
of massive galaxies. Surveys over
the full set of optical/near-IR passbands from
$U$ through $K$ are essential to obtain representative samples
of massive galaxies.

The main uncertainty in this analysis  is the
reliance on photometric redshifts. We estimated the effect of
this uncertainty by randomly perturbing the redshifts
using a Gaussian distribution with dispersion $\Delta z/(1+z)=0.12$,
and repeating the selection and analysis. Despite significant
migration of galaxies in and out of the $2<z<3$ redshift range the
values in Table 1 change by only $\sim 10$\,\%.
We note, however, that
there may be subtle systematic biases which can
have significant effects, in particular on the derived
masses (see, e.g., Shapley et al.\ 2005).
Comprehensive tests of the techniques employed here and in other
studies of infra-red selected samples (e.g., Dickinson et al.\ 2003,
Rudnick et al.\ 2003) are urgently needed, and
will become feasible with the introduction of multi-object
near-IR spectrographs on 8m class telescopes.

\begin{acknowledgements}
We thank the anonymous referee for insightful comments which improved
the manuscript significantly.
PGvD acknowledges support from NSF CAREER AST-0449678. DM is supported
by NASA LTSA NNG04GE12G. EG is supported by
NSF Fellowship AST-0201667. ST is partly supported by the
Danish Natural Research Council.
\end{acknowledgements}

\bibliography{}

\end{document}